\documentclass[a4paper,10pt,twoside,twocolumn]{article}
\usepackage{ercoftac_2012}
\usepackage{etoolbox}
\usepackage[utf8]{inputenc}

\begin{document}

\title{{\vspace{-0.0cm}\sc{Enhancing the predictive capabilities for high P/T fuel sprays; non-ideal thermodynamic modelling using PC-SAFT}}}
\author{Phoevos Koukouvinis$^{1,2}$, Alvaro Vidal-Roncero$^1$ \\  Carlos Rodriguez$^1$, Manolis Gavaises$^1$, Lyle Pickett$^2$\vspace{0.3cm}\\
\small{\it{$^1$SMCSE, City University London, Northampton Square EC1V 0HB, United Kingdom}}\\
\small{\it{$^2$Combustion Research Facility, Sandia National Laboratory, 7011 East Ave, Livermore, CA 94550, US}}\\}
\date{}
\maketitle
\thispagestyle{fancyplain}
\vspace{0.5cm}

\section*{Abstract}
The present work aims to investigate the complex phenomena occurring during high-pressure/high-temperature fuel injection of the Engine Combustion Network (ECN) Spray-A case. While commonly in the literature transcritical mixing cases are approached using traditional cubic equation-of-state models, such models can prove insufficient in the accurate prediction of liquid density and speed of sound. The purpose of the present investigation is to employ a general tabulated approach which can be applied to any type of thermodynamic closure. At the same time, a more advanced model based on the Perturbed-Chain Statistical Associating Fluid Theory (PC-SAFT) is employed to create the thermodynamic table, as it is proven superior to the traditional cubic models, while also having the capacity of predicting Vapor-Liquid-Equilibrium. The model has been used for a combination of dodecane and nitrogen mixing, corresponding to the well known Spray-A conditions. Vapor penetration and mixing both in terms of temperature and mass fraction are found in agreement to experiments, within the experimental errors. Also, the thermodynamic states correspond well with the adiabatic isobaric-mixing curve, demonstrating the energy-conservative nature of the approach.

\section{Introduction}
The operation of modern Internal Combustion Engines (ICEs) involves injection and mixing of fuel with oxidiser and, consequently, combustion, as the core of the engine operation; indeed, the processes of fuel atomisation, evaporation, mixing with the gas and combustion greatly affect engine efficiency and emissions \cite{reitz2019}, hence are extensively studied both by industry and academic/research institutes. A particular complexity of the operation of modern engines, is the extreme temperature/pressure range that the fuel undergoes from the fuel piping, then through the injector and finally in the combustion chamber/engine cylinder. Indicatively, for modern Diesel engines, a pressure variation from more than 2000bar (at common rail, note that modern systems may even reach 3000bar) to effectively 0bar (cavitation regions in the fuel injector) and a temperature variation from ~363K (high pressure side of the fuel pump) to more than 1000K (engine cylinder) are expected. At such pressure/temperature ranges fuel properties cannot be assumed constant. Indicatively, diesel density may change by ~14\% over a range of 0-2000bar, at 363K; for the same range viscosity can change over 200\% \cite{Kolev2012}. Such variations have severe thermodynamic implications in the behaviour of the fuel. In particular, the expansion of the fuel as pressure drops induces cooling due to the Joule-Thomson effect. On the other hand, the immense shear stresses the fuel is subjected into, cause strong heating. Previous works have identified the dependence of induced heating or cooling on the fuel injector discharge coefficient, which is a measure of the efficiency of the fuel injector. \\
Additional complexities that take place along property variation, involve phase change mechanisms, such as cavitation at sharp geometric features inside the injector or evaporation and mixing with the oxidiser. Also, as the downstream conditions of the injector may exceed the critical point of the fuel/gas, departure from classical atomisation may occur, giving its place to transcritical mixing. Indeed, evidence exists that, at elevated pressures and temperatures, surface tension effects diminish and the fuel/gas mixing is dominated by diffusive mixing \cite{manin2014}. It is notable, that the aforementioned effects  are not solely related to Diesel engines, as similar transcritical operation is also relevant to gasoline engines \cite{yang2016}, high-pressure gas turbines \cite{dahms2013} and rocket engines \cite{chehroudi2012}. 
The previous description of phenomena during fuel injection provides a brief overview of the complexities that necessitate the use of accurate real-fluid modelling, as it is necessary to capture phase transitions, subcritical/transcritical/supercritical mixing and temperature changes due to fuel expansion. Traditionally, models capable of describing such phenomena are cubic Equations of State (EoS); a representative example is the Peng Robinson EoS\cite{peng1976}. Despite their widespread use, cubic EoS are known to suffer from deficiencies, as they commonly underpredict liquid density and overpredict speed of sound (the interested reader is addressed to \cite{perez2017,kikic2001}). Improvements, aiming mainly to address the liquid density exist; e.g. the generalised Redlich-Kwong-Peng-Robinson EoS\cite{cismondi2005}, or volume translated methods\cite{nazarzadeh2013}, however, inaccuracies or inconsistencies are still present. On this aspect, there are more complex models, which however require extensive calibration based on experimental data (see e.g. \cite{lemmon2004}). An attractive alternative, is the Perturbed Chain Statistical Associating Fluid Theory (or PC-SAFT) model, which has a higher accuracy comparing to cubic EoS in predicting thermodynamic and transport properties, both for pure components and mixtures, requiring minimal input of just three molecular characteristics of the component to be modelled.    \\
In the present work, the PC-SAFT model is used to predict properties of a hydrocarbon and nitrogen mixture, aiming to replicate a well-known benchmark case in the area of high pressure/high temperature fuel injection, the Spray-A case from the Engine Combustion Network (ECN). In the next sections, the mathematical and thermodynamic models are presented, followed by the geometry of the Spray-A case and the case set-up. A detailed comparison between the numerical simulations and the experimental data is made, examining the mass fraction and temperature distributions in the radial and axial direction of the spray.

\section{Mathematical model}
The simulations are carried out by assuming a diffuse interface approach, under a homogeneous mixture assumption under mechanical and thermal equilibrium\cite{prosperetti2000} ; all fluids involved share the same velocity, pressure and temperature fields. Hence, the model consists of four Partial Differential Equations, plus the thermodynamic closure. It is highlighted that the term mixture is used to denote multicomponent (multiple materials) mixture, rather than multiple phases.
The governing equations are provided below: \\
-Mixture mass conservation equation:
\begin{eqnarray}
\frac{\partial \rho}{\partial t} + \nabla \cdot (\rho \textbf{u}  ) = 0
\label{eq:1}
\end{eqnarray}
where \( \rho \) is the mixture density and \( \textbf{u} \) is the velocity vector field.  \\
- Dodecane mass fraction, \( \textit{y\textsubscript{C12}}\), transport equation:
\begin{eqnarray}
\frac{\partial \rho y\textsubscript{C12}}{\partial t} + \nabla \cdot (\rho \textbf{u}  y\textsubscript{C12} ) = -\nabla \cdot \textbf{J}
\label{eq:2}
\end{eqnarray}
where \textbf{J} is the mass diffusion flux defined as: \\
\begin{eqnarray}
\textbf{J} = - \left(\rho D\textsubscript{m}+ \frac{\mu\textsubscript{t}}{Sc\textsubscript{t}}\right)\nabla y\textsubscript{C12} - D\textsubscript{T} \frac{\nabla T}{T}
\label{eq:3}
\end{eqnarray} \\
\textit{D}\textsubscript{m} and \textit{D}\textsubscript{T} are the mass and temperature diffusion coefficients respectively and \textit{Sc}\textsubscript{t} is the turbulent Schmidt number, which plays an important role, as the simulations are performed with a RANS turbulence model. The value of turbulent Schmidt number used here is 0.5. 
The diffusion coefficients are calculated using kinetic theory, as:
\begin{eqnarray}
 D\textsubscript{m} = 0.00188\frac{ \left[  T^3\left( \frac{1}{MW\textsubscript{C12}} + \frac{1}{MW\textsubscript{N2}} \right)  \right]^{1/2}}{p\sigma^2\Omega \textsubscript{D}}
\label{eq:4}
\end{eqnarray} \\
were \( \Omega \textsubscript{D}\) is the diffusion collision integral, which can be found in tabulated form or analytic expressions (see \cite {Poling2000}) and is a function of  \( T* = \frac{T}{\epsilon/k\textsubscript{B}}\). The temperature diffusion coefficient is calculated based on the multi-component approximation of \cite {Kuo2012}. Quadratic Upwind Interpolation For Convective Kinematics (QUICK) has been used for the transport equation of dodecane. The parameters needed for calculating the diffusion coefficients are presented in \ref{tab:1} (from \cite {Tahery2007}). 
\begin{table}[h]
\begin{center}
\caption{Kinetic theory parameters for the components examined}
\label{tab:1}
\begin{tabular}{lll}
    & \( \sigma  ({\AA}) \)    & \( \epsilon/k\textsubscript{B} \) (K)      \\
C\textsubscript{16} & 8.89 & 764.03 \\
C\textsubscript{14} & 8.3  & 701.92 \\
C\textsubscript{12} & 7.58 & 622.51 \\
N\textsubscript{2}  & 3.4  & 102.12
\end{tabular}
\end{center}
\end{table}
The values used for estimating \textit{T}* and \textit{D}\textsubscript{m} are arithmetic and geometric averages of the involved components, respectively.  \\
- Mixture Momentum equation:
\begin{eqnarray}
\frac{\partial \rho \textbf{u}}{\partial t} + \nabla \cdot (\rho \textbf{u} \otimes  \textbf{u} )= -\nabla p + \nabla \cdot \mathbf{\tau}
\label{eq:5}
\end{eqnarray}
where \textit{p} stands for the pressure and \( \tau \) corresponds to the stress tensor  \( \tau=\mu\textsubscript{eff} [\nabla u +(\nabla u)^T ] \)  , with \( \mu\textsubscript{eff} \) the sum of laminar,  \( \mu\), and turbulent,  \( \mu\textsubscript{t} \), dynamic viscosity. The momentum equation is resolved using a Second Order Upwind \cite {Versteeg2007} scheme, to minimise numerical diffusion, while also maintaining stability. \\
- The mixture energy equation:
\begin{eqnarray}
\frac{\partial \rho E}{\partial t} + \nabla \cdot (\textbf{u} (\rho E+p) ) &= & \nabla \cdot (\lambda \textsubscript{eff} \nabla T ) + \nabla \cdot (\mathbf{\tau} \cdot \textbf{u}) \nonumber \\
&&   + \nabla \cdot (h\textbf{J})
\label{eq:6}
\end{eqnarray}
where \( E \) is the total energy, defined as internal energy plus the kinetic energy, or \( E=h-\frac{p}{\rho}+\frac{1}{2}\textbf{u}^2 \) where \( h \) is the enthalpy, provided as a function of pressure and temperature (see section of Thermodynamic Properties). The total effective thermal conductivity, \( \lambda \textsubscript{eff} \), is equal to the thermal conductivity, \( \lambda \), being a function of thermodynamic conditions \(  p,  T  \) and composition, \( y\textsubscript{C12}  \), plus the turbulent contribution, as follows:
\begin{eqnarray}
\lambda \textsubscript{eff}=\lambda + \frac{c\textsubscript{p} \mu\textsubscript{t} }{Pr\textsubscript{t}}
\label{eq:7}
\end{eqnarray}
where \( c\textsubscript{p} \) is the heat capacity of the mixture and \( Pr\textsubscript{t} \) is the turbulent Prandtl number, assumed equal to 0.85, based on the average value obtained from multiple experiments using different materials \cite {Kays2005}. The energy equation is discretized using a Second Order Upwind scheme \cite {Versteeg2007}.
Turbulence is handled with the standard k-\( \epsilon \)  model, its coefficients, in particular Schmidt number and \( C \textsubscript{$\epsilon$1} \), tuned based on previous studies \cite {Wei2017}, to values of 0.5 and 1.52 respectively. 
\begin{eqnarray}
\frac{\partial \rho k}{\partial t} + \nabla \cdot (\rho \textbf{u} k ) &= & \nabla \cdot \left[ \left( \mu+\frac{\mu\textsubscript{t}}{\sigma\textsubscript{k}}   \right) \nabla k \right] \nonumber \\
&& +G\textsubscript{k} - \rho\epsilon-Y\textsubscript{M}
\label{eq:8}
\end{eqnarray}
\begin{eqnarray}
\frac{\partial \rho \epsilon}{\partial t} + \nabla \cdot (\rho \textbf{u} \epsilon ) &= & \nabla \cdot \left[ \left( \mu+\frac{\mu\textsubscript{t}}{\sigma\textsubscript{$\epsilon$}}   \right) \nabla \epsilon \right] \nonumber \\
&& + C \textsubscript{$\epsilon$1}\frac{\epsilon}{k}G\textsubscript{k} - C \textsubscript{$\epsilon$2}\rho\frac{\epsilon^2}{k}
\label{eq:9}
\end{eqnarray}
where \( G\textsubscript{k} \) is the turbulence generation term and \( Y\textsubscript{M} \)  is the turbulence dilation term \cite {Sarkar1991}.

\section{Thermodynamic model}
\begin{figure}[h]
 \includegraphics[width=0.45\textwidth]{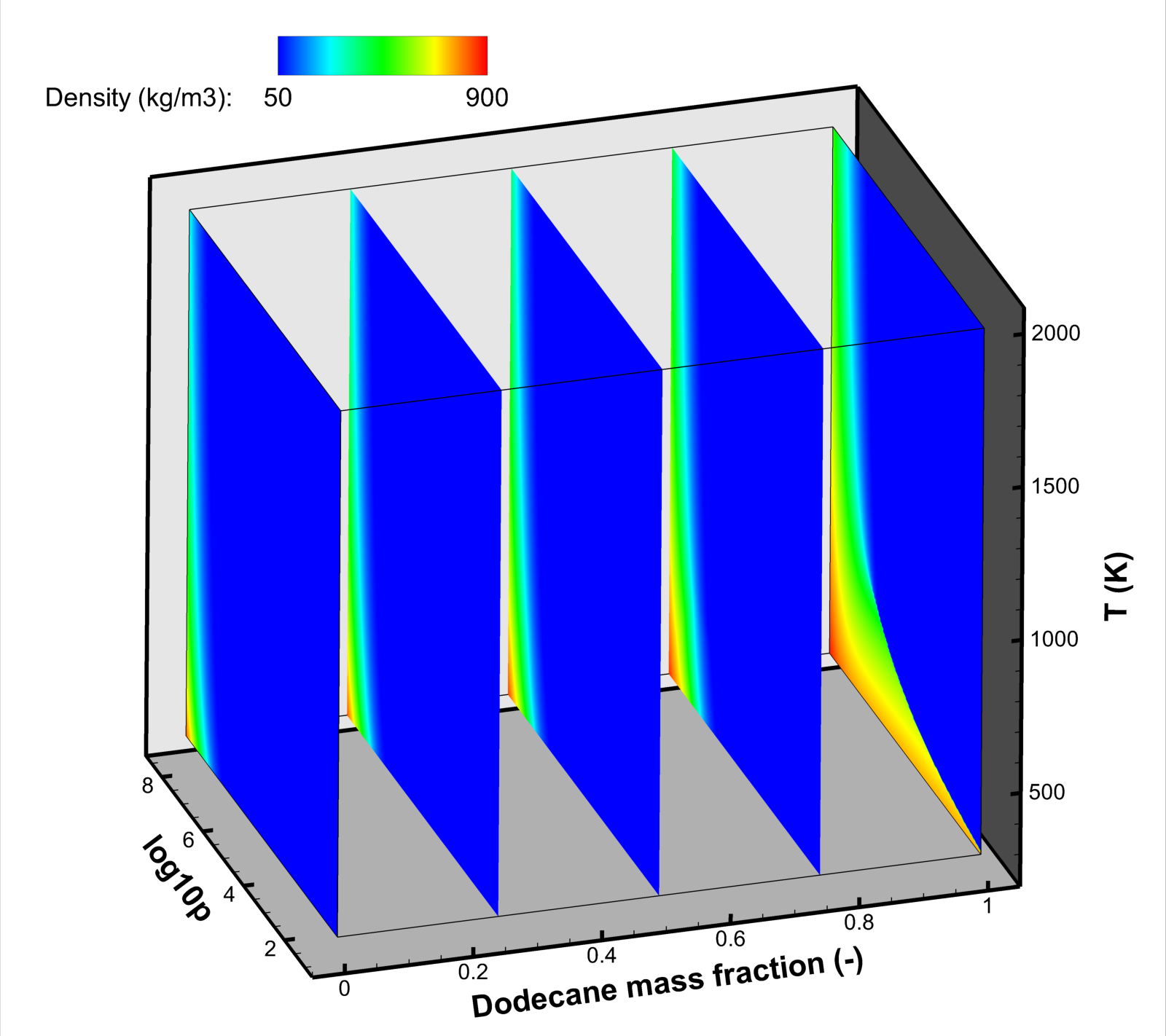}
 \caption{The dodecane(C\textsubscript{12})- nitrogen phase diagram, used for the CFD solver. Indicative slices are shown for different C\textsubscript{12} mass fractions: 0 (pure nitrogen), 0.25, 0.5, 0.75, 1.0 (pure dodecane). Coloring is according to the mixture density.}
 \label{fig:01}
\end{figure}

Accurate modelling of fuel/gas properties is done using the Perturbed Chain Statistical Associating Fluid Theory (PC-SAFT) EoS, which is a theoretically derived model, based on perturbation theory, that splits the intermolecular potential energy of the fluid into a reference term accounting for repulsive interactions and a perturbation term accounting for attractive interactions. The reference fluid is composed of spherical segments comprising a hard sphere fluid that then forms molecular chains to create the hard-chain fluid. The attractive interactions, perturbations to the reference system, are accounted for with the dispersion term. Intermolecular interaction terms accounting for segment self- or cross-associations are ignored. Hence, each component is characterized by three pure component parameters, which are a temperature-independent segment diameter, \( \sigma \), a segment interaction energy,  \( \epsilon/k \), and a number of segments per molecule, \( m \); detailed databases for these parameters exist for non-associating fluids, such as hydrocarbons or gases, see \cite{gross2001, polishuk2014}.  \\
The PC-SAFT model aims to construct an expression for the residual Helmholtz energy \( a\textsubscript{res} \), which is the sum of a hard-chain term and a dispersion term; the detailed expression of these terms is rather lengthy, hence the interested reader is addressed to the original model publication \cite{gross2001}. Once the residual Helmholtz energy is obtained, all thermodynamic properties can be defined as functions of that expression. The transport properties are estimated based on the residual entropy scaling method, following L{\"o}tgering-Lin and Gross  \cite{lin2015} for dynamic viscosity and  Hopp and Gross \cite{hopp2017} for thermal conductivity. \\ 
Identification of Vapor Liquid Equilibrium (VLE) is done through the the minimisation of the molar Helmholtz Free energy, defined in terms of density, temperature and composition.This optimization problem is solved via a combination of the successive substitution iteration (SSI) and the Newton minimization method with a two-step line-search procedure, and the positive definiteness of the Hessian is guaranteed by a modified Cholesky decomposition. The algorithm consists of two stages: first, the mixture is assumed to be in a single-phase state and its stability is assessed via the minimization of the Tangent Plane Distance (TPD) \cite{yang2020}. The stability is tested by purposely dividing the homogeneous mixture in two phases, one of them in an infinitesimal amount and called 'trial phase'. For any feasible two-phase mixture, if a decrease in the Helmholtz free energy is not achieved, then the mixture is stable. In case the minimum of the TPD is found to be negative, the mixture is considered unstable and a second stage of phase splitting takes place consisting on the search for the global minimum of the Helmholtz Free Energy. As a result, the pressure of the fluid and the compositions of both the liquid and vapor phases are calculated. \\
All the aforementioned process for calculating properties and identifying VLE is done as a precursor step of the simulation, to construct a structured table, where all properties are expressed in terms of the decimal logarithm of pressure (\(log\textsubscript{10}p \)), temperature and mass fraction. The CFD solver uses this table and performs linear interpolation to identify all thermodynamic properties required during the simulation. In the present work, the thermodynamic table had a resolution of 100 x 400 x 101 corresponding to \(log\textsubscript{10}p \), \(T\), \(y\) intervals, for a range of \(p\):[10Pa - 2500bar] x \(T\):[280 - 2000K] x \(y\):[0-1] respectively. The structure of the table enables very fast searching and reasonable accuracy in the property representation; validation of the method is in the previous work of the authors, see \cite{koukouvinis2020}. An indicative representation of the tabulated EoS used in terms of density is shown in \ref{fig:01}.    

\section{Simulations}
\subsection{Case set-up}
\begin{figure}[h]
 \includegraphics[width=0.45\textwidth]{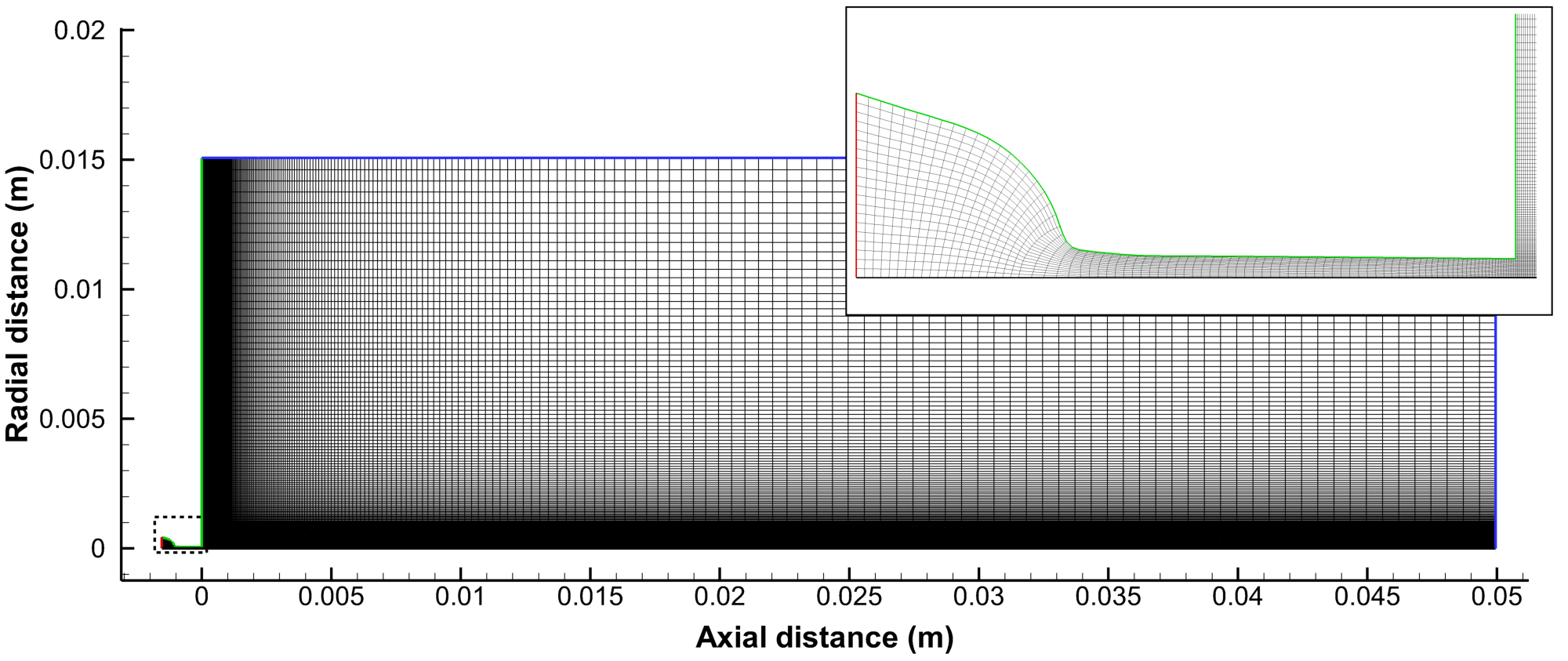}
 \caption{Axis-symmetric geometry of the Spray-A injector used, along with the downstream injection chamber and computational mesh. Boundary conditions are also shown: fixed mass flow, composition and temperature (fuel inlet, red), walls (green), fixed pressure outflow (blue). The magnified insert corresponds to the dashed region, showing the mesh inside the injector. X-axis is the axis of symmetry.}
 \label{fig:02}
\end{figure}

The Spray-A injector is a single hole, tapered (k-factor=1.5) Bosch solenoid-activated injector, with nominal orifice diameter \(d\textsubscript{out} \)$\sim$90\(\mu\)m and orifice length of ~1mm. The geometry of the Spray-A injector used is based on the published geometry from the ECN website. The simulations to be presented, are treated as 2D axis-symmetric, using the published radial profile over the injector axis \cite{sprayA}. Apart from the injector, the computational domain is extended downstream 50mm in the axial and 15mm in the radial directions (or by $\sim$\(56d\textsubscript{out}\) and $\sim$\(17d\textsubscript{out}\) respectively), to include part of the spray chamber over which measurements are available. The computational mesh consists of 100k finite volumes, is purely quadrilateral and is refined in the injector orifice and near the exit of the injector, as shown in \ref{fig:02}; the resolution inside the orifice is 2\(\mu\)m (resulting to a maximum y+ of $\sim$50). Boundary conditions correspond to the Spray-A case for injection pressure of 1500bar, 363K to stagnant nitrogen atmosphere at (a) 60bar and 900K and (b) 50bar and 1100K\cite{ECNref}. To emulate the needle motion, a mass flow profile is introduced upstream the injector, based on the online tool from CMT-Motores T\'ermicos \cite{cmt}. Apart from the mass flow rate, temperature of the fuel is fixed at 363K and the composition to pure dodecane. At the far-field, fixed pressure is imposed equal to the  chamber pressure downstream the injector. Temperature and composition are imposed as zero gradient at farfield, unless backflow is found; in that case fixed composition and temperature are imposed based on the chamber conditions. Nozzle walls are considered as adiabatic, as the injector is kept at the same temperature as the fuel. 

\subsection{Results}
\begin{figure}[h]
 \includegraphics[width=0.45\textwidth]{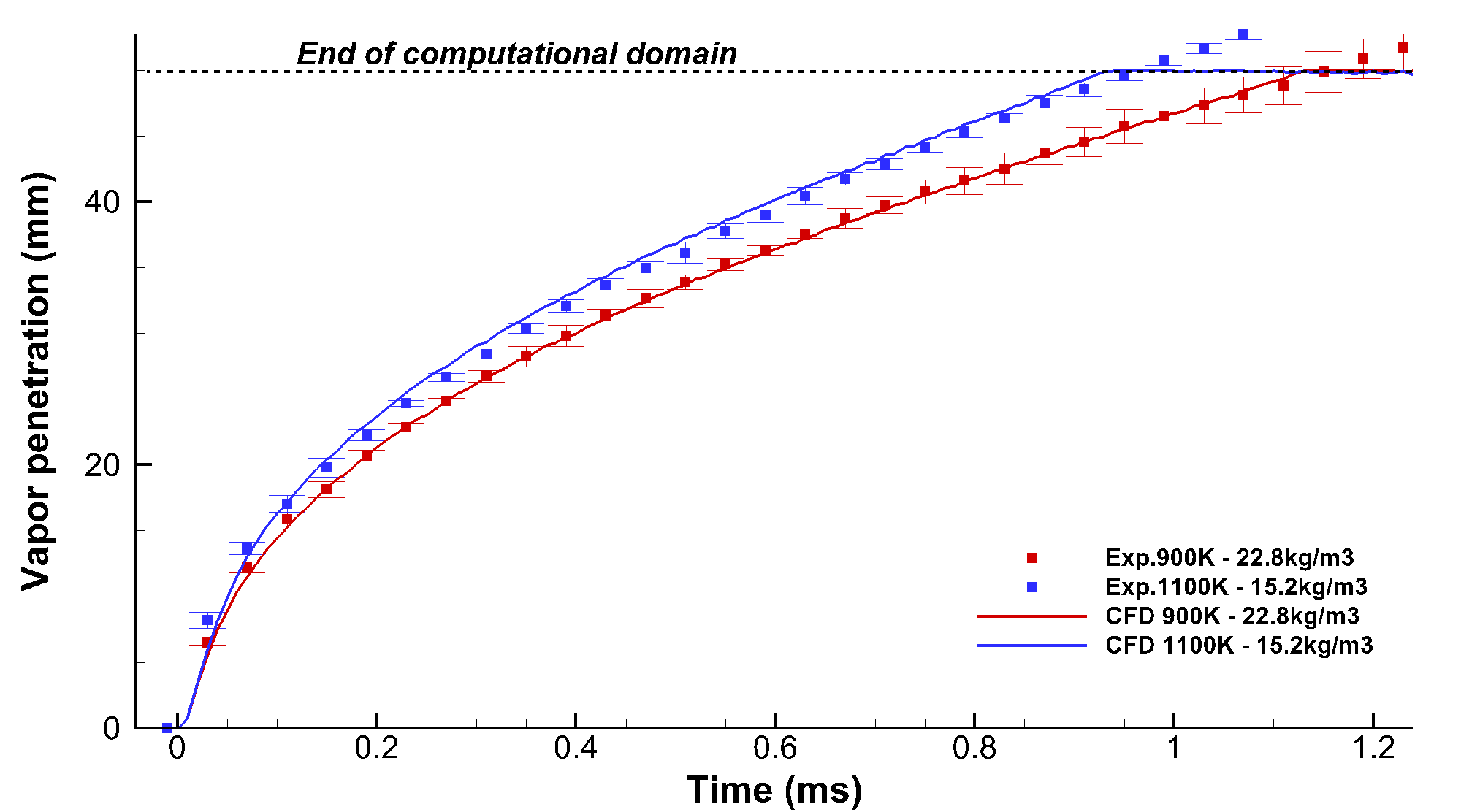}
 \caption{Vapor penetration over time for two downstream conditions; (red) 900K and 60bar (\(22.8kg/m^3\)), (blue) 1100K and 50bar (\(15.2kg/m^3\)). Symbols correspond to experiments, continuous line to simulations.}
 \label{fig:03}
\end{figure}

\begin{figure}[h]
 \includegraphics[width=0.45\textwidth]{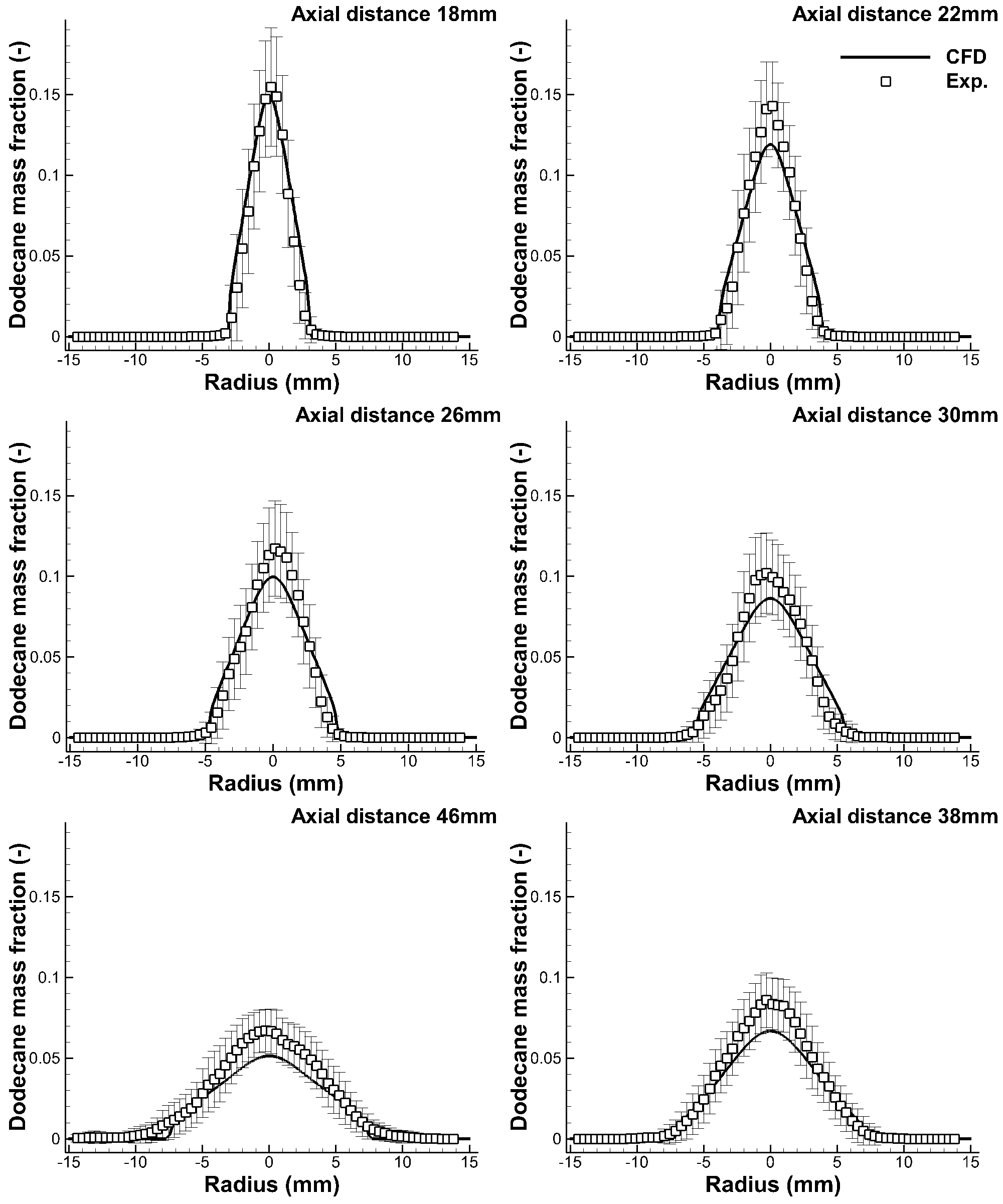}
 \caption{Radial distribution of dodecane mass fraction at different positions from the injector, injection to nitrogen,  900K and 60bar. }
 \label{fig:04}
\end{figure}

\begin{figure}[h]
 \includegraphics[width=0.45\textwidth]{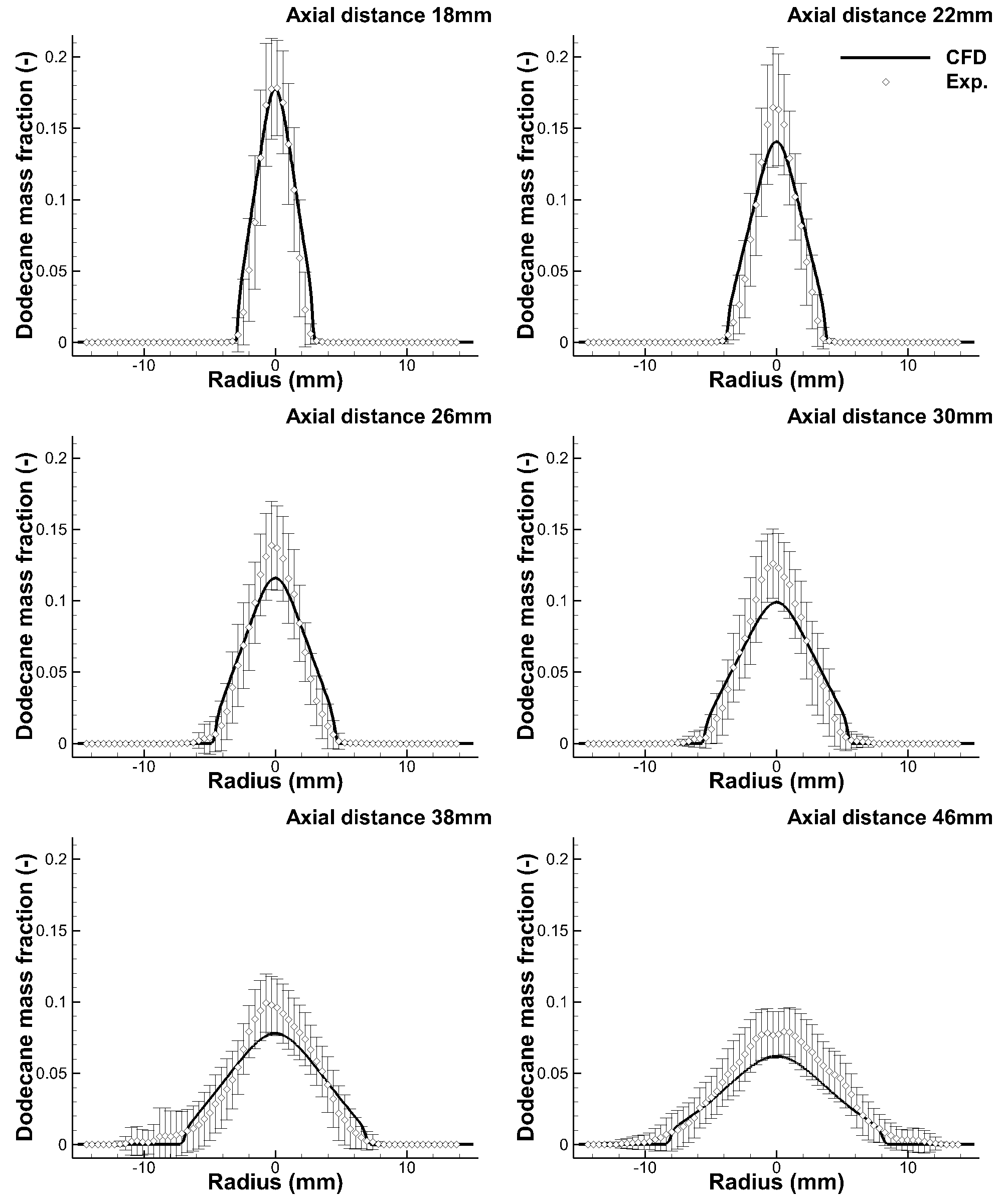}
 \caption{Radial distribution of dodecane mass fraction at different positions from the injector, injection to nitrogen, 1100K and 50bar.}
 \label{fig:05}
\end{figure}

\begin{figure}[h]
 \includegraphics[width=0.45\textwidth]{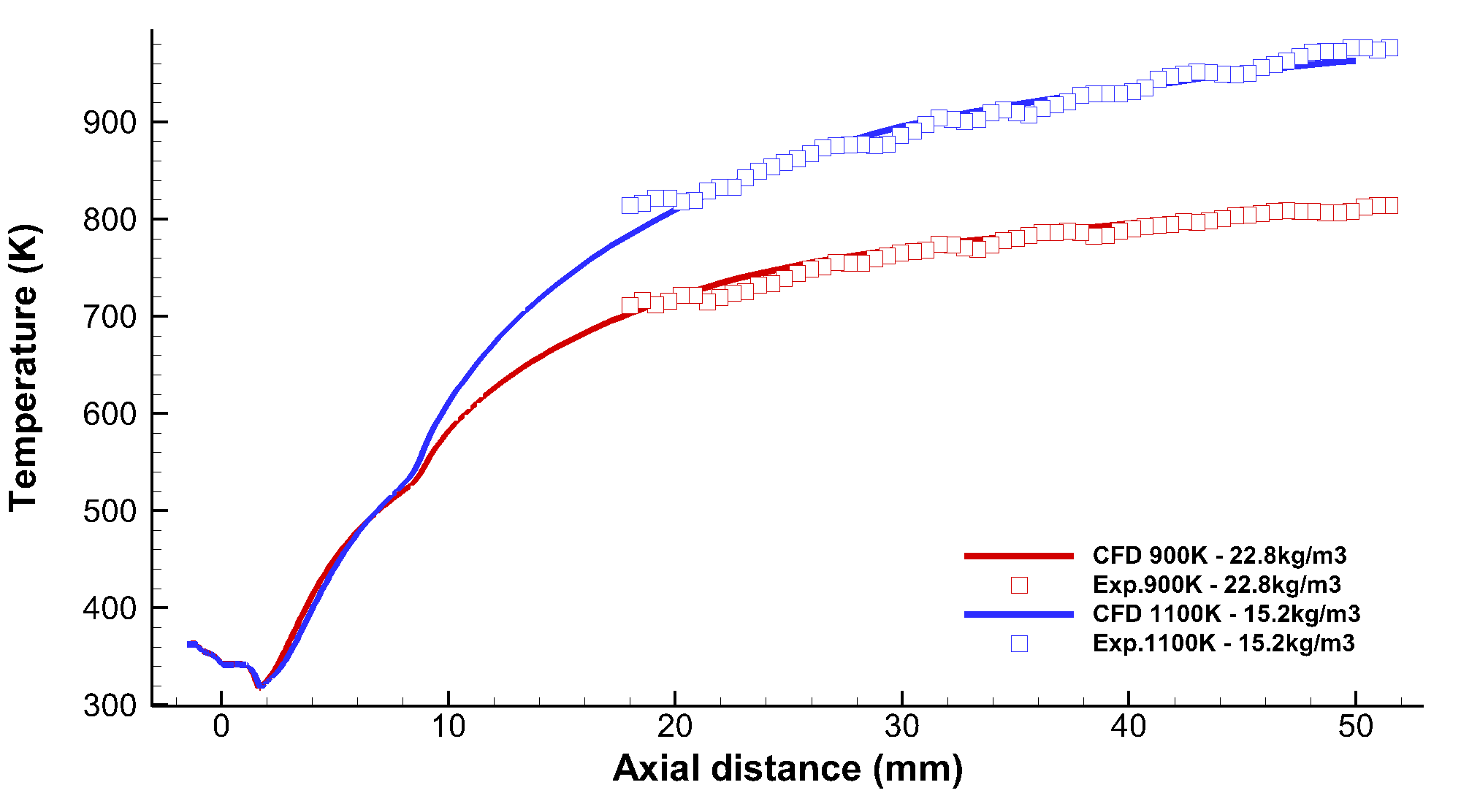}
 \caption{Temperature distribution, along the jet axis, for two downstream conditions; (red) 900K and 60bar (\(22.8kg/m^3\)), (blue) 1100K and 50bar (\(15.2kg/m^3\)). }
 \label{fig:06}
\end{figure}

In  \ref{fig:03} the temporal evolution of the vapor penetration of the dodecane jet is shown, for the two ambient conditions examined. Vapor penetration was identified by defining the isosurface of mass fraction equal to 0.1\%, following ECN guidelines, and calculating its maximum axial coordinate for every time instant. As expected, penetration is faster at 1100K and 50bar, since the ambient density is much lower (by $\sim$50\%) at this condition. The agreement with experimental results is within the experimental uncertainty, hence is considered satisfactory. Further to instantaneous vapor pentration, the overal distribution of dodecane mass fraction downstream the injector is within good agreement with experiments, as shown in \ref{fig:04} for 900K/60bar and in \ref{fig:05} for 1100K/50bar. Apart from mass fraction, a comparison of the axial temperature distribution is shown in  \ref{fig:06}; the agreement between numerical redictions and the temperature estimations from the experiment are very close. Note that in the region of -2 to 2mm, there is temperature reduction, due to the depressurization of the compressed liquid and mixing with nitrogen.

\begin{figure}[h]
 \includegraphics[width=0.45\textwidth]{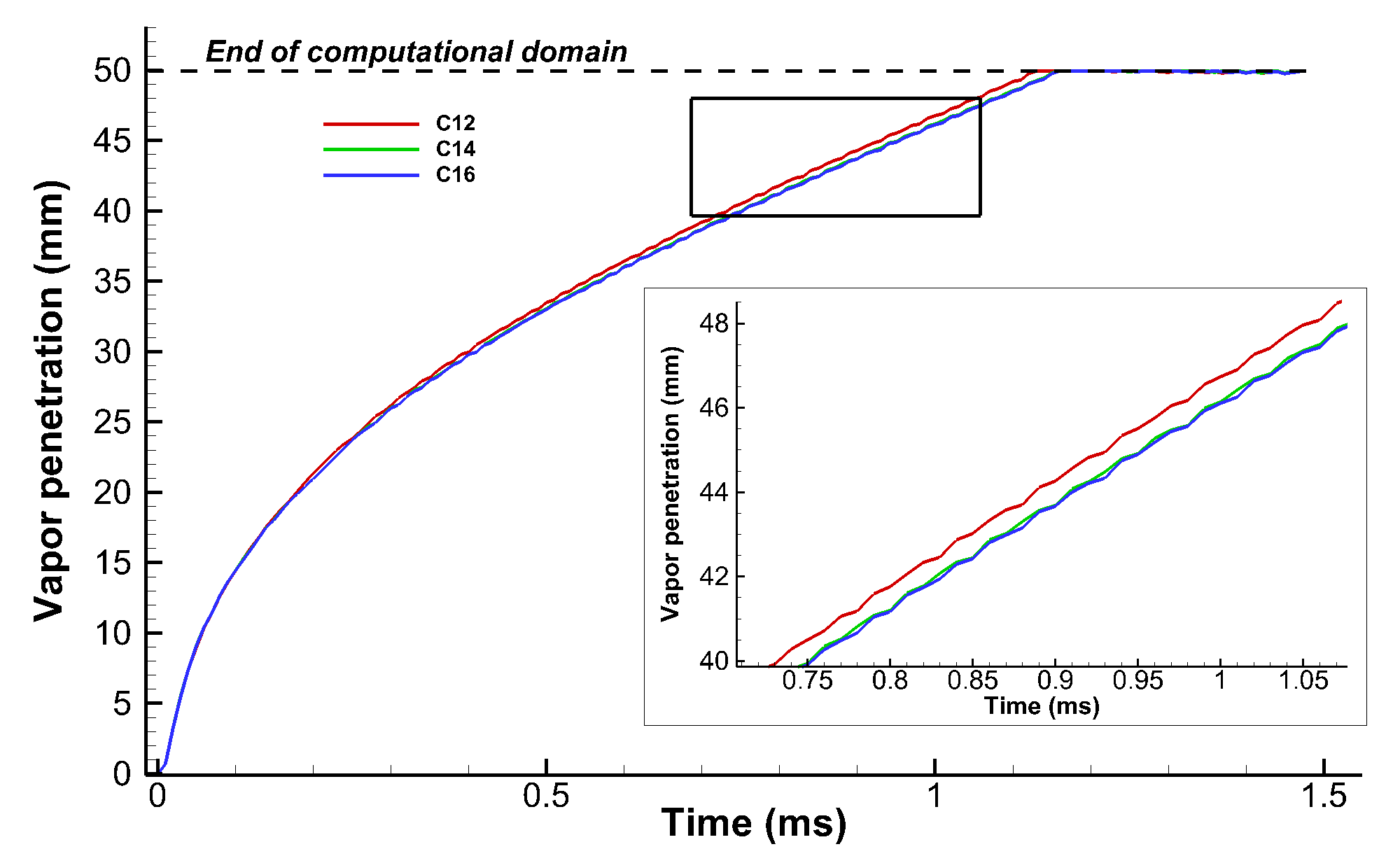}
 \caption{Vapor penetration over time for different fuels, dodecane, tetradecane and hexadecane injected to nitrogen, at 900K and 60bar. }
 \label{fig:07}
\end{figure}

The predictive capability of the PC-SAFT model, is demonstrated when comparing the vapor penetration of different hydrocarbons. An example here is in \ref{fig:07}, for dodecane, tetradecane and hexadecane, all modeled with the same methodology. As shown, the dodecane has slightly higher penetration rate over the other examined hydrocarbons, of which tetradecane is marginally faster than hexadecane. This is expected to happen for two reasons; (1) dodecane density within the injector ranges between \(714-786kg/m^3\), whereas for tetradecane \(732-803kg/m^3\) and hexadecane \(740-809kg/m^3\). Hence, the larger the density, the lower the jet velocity, thus slightly slower penetration. (2) for the three examined hydrocarbons, molecular weight correlates with viscosity. Indeed, at the injector outlet, dodecane viscosity is 0.654mPa.s, tetradecane is 0.795mPa.s and hexadecane is 1.016mPa.s, hence it is expected that the higher the viscosity, the lower the jet velocity due to viscous losses. \\

\begin{figure}[h]
 \includegraphics[width=0.45\textwidth]{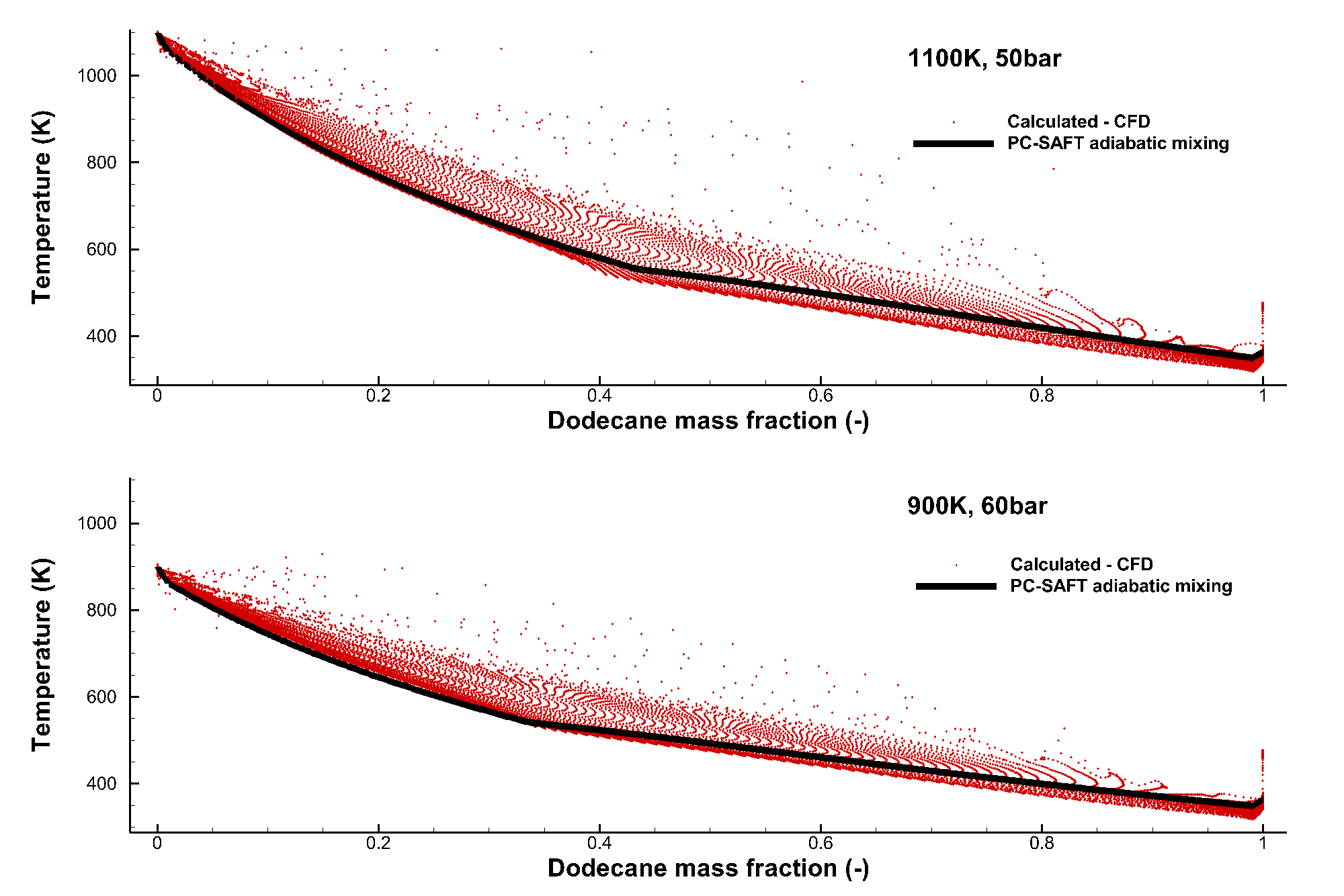}
 \caption{Adiabatic mixing curves, estimated using PC-SAFT, superimposed with the CFD states predicted by the solver, at (up) 1100K/50bar (down) 900K/60bar. }
 \label{fig:08}
\end{figure}

A further validation of the simulations is performed by comparing the CFD solver states with the isobaric, adiabatic mixing curves. For PC-SAFT mixing curves are found directly by using the tabulated EoS, shown in \ref{fig:01}, assuming the initial temperature of the two components at the downstream pressure, calculating enthalpy, then by calculating all the intermediate mixing states assuming isenthalpic mixing at constant pressure (i.e. \(h\textsubscript{mix}= y\textsubscript{C12}h\textsubscript{C12}+y\textsubscript{N2}h\textsubscript{N2}\) ) and finally inverting enthalpy to temperature; it serves as a representation of the cooling due to fuel/gas mixing. Also it serves as a validation of the solver, as roughly the CFD states should follow the of the mixing curve. As shown in \ref{fig:08} the CFD solver states (represented with red dots) follow the trend of the adiabatic, isobaric PC-SAFT mixing curve for both cases. Some scatter is observed, as the injector has been included in the orifice, hence the fuel is not injected at a constant temperature; inded, fuel near the center of the orifice is colder due to the Joule-Thomson effect, whereas fuel near the orifice walls will be hotter, due to viscous effects. This can be clearly demonstrated from \ref{fig:08}, as the temperatures of the pure dodecane component range from $\sim$341K (which corresponds to liquid core temperature; this is the lowest temperature that can be achieved by isentropic expansion of dodecane from 1500bar and 363K)  to $\sim$476K near the walls. It is also notable, that the lowest temperature (of $\sim$318K) is achieved further downstream the injector orifice outlet, due to dodecane mixing with nitrogen, for a dodecane mass fraction of 99\%.

\section{Conclusions}
In the present work, a numerical model is presented for simulating the complex interaction of fuel mixing with gas, for conditions relevant to modern ICEs. The model is based on accurate thermodynamic modelling using PC-SAFT, to provide properties of pure components and their mixtures, over a wide range of conditions, for which idealised simplifications are grossly insufficient. The model predictions are accurate against experiments, both in terms of instantaneous vapor penetration and the average distributions of mass fractions and temperature. In general, the further from the injector, the agreement becomes worse, though even at the furthest distance examined, at 46mm, is within the experimental uncertainty. Probably this is related to the inherent deficiencies of RANS turbulence models; LES modelling would be expected to provide a much better mass fraction distribution, without the modelling assumptions of the k-\( \epsilon \) model. The difference in computational cost between the present approach and an LES study has to be stressed though; indicatively, the results presented in this work are obtained using a regular workstation within 10 hours for a single examined case, making the method attractive for parametric investigations. Furthermore, another attractive feature of the PC-SAFT model is the possibility to investigate the mixing characteristics of different hydrocarbons and even hydrocarbon mixtures (fuel surrogates or real fuels) with relative ease and minimum experimental calibration. This enables to examine what-if scenarios, for different fuels, which will be investigated in the future.

\section*{Acknowledgments}
This work has received funding from the European Union's Horizon 2020 research and innovation programme under the Marie Sklodowska-Curie grant agreement No 748784. The primary author would also like to thank Sandia National Laboratories, the US Department of Energy (DoE) and the staff at Combustion Research Facility for hosting him. Sandia National Laboratories is a multi-mission laboratory managed and operated by National Technology and Engineering Solutions of Sandia, LLC., a wholly owned subsidiary of Honeywell International, Inc., for the U.S. Department of Energy's National Nuclear Security Administration under contract DE-NA0003525.

%Bibliography section! 
\bibliography{references} % corresponds to references.bib file name
\bibliographystyle{ieeetr}
\end{document}